\documentclass[twocolumn,amsmath,amssymb,prx,longbibliography]{revtex4-1}

\usepackage{graphicx}
\usepackage{dcolumn}
\usepackage{bm}
\usepackage{color}
\usepackage{notes2bib}

\begin{document}



\title{Highly Efficient Optical Pumping  of Spin Defects in Silicon Carbide for Stimulated Microwave Emission}

\author{M.~Fischer$^{1}$}
\author{A.~Sperlich$^{1}$}
\author{H.~Kraus$^{1}$}
\author{T.~Ohshima$^{2}$}
\author{G.~V.~Astakhov$^{1}$}
\email[E-mail:~]{astakhov@physik.uni-wuerzburg.de}
\author{V.~Dyakonov$^{1,3}$}
\email[E-mail:~]{dyakonov@physik.uni-wuerzburg.de}

\affiliation{$^1$Experimental Physics VI, Julius-Maximilian University of W\"{u}rzburg, 97074 W\"{u}rzburg, Germany \\
$^2$National Institutes for Quantum and Radiological Science and Technology, Takasaki, Gunma 370-1292, Japan \\
$^3$Bavarian Center for Applied Energy Research (ZAE Bayern), 97074 W\"{u}rzburg, Germany}

\begin{abstract}
We investigate the pump efficiency  of silicon vacancy-related spins in silicon carbide. For a crystal inserted into a microwave cavity with a resonance frequency of  9.4~GHz, the spin population inversion factor of 75 with the saturation optical pump power of about 350~mW is achieved at room temperature. At cryogenic temperature,  the pump efficiency drastically increases, owing to an exceptionally long spin-lattice relaxation time exceeding one minute. Based on the experimental results, we find realistic conditions under which a silicon carbide maser can operate in continuous-wave mode and serve as a quantum microwave amplifier. \end{abstract}

 
\date{\today}

\maketitle

\section{Introduction}

Highly-sensitive microwave (MW) detectors are in the heart of many modern technologies, such as long-distance communication, 
time keeping, remote sensing, magnetic resonance tomography and quantum information processing. Masers, exploiting a pure quantum phenomenon---stimulated MW emission due to a population inversion of spin states---can amplify vanishingly  weak MW signals with adding only marginal noise. Since invention in the 50's, most of the masers operate in extreme environments \cite{siegman1964microwave}. A breakthrough was a demonstration of a pentacene maser in tabletop experiments under ambient conditions \cite{Oxborrow:2012dd}. However, only pulsed mode with a relatively high optical pump power of $70 \, \mathrm{W}$ has been realized so far \cite{Breeze:2015ea, Salvadori:2017hv}. 

An alternative approach to achieve continuous-wave (cw) masing is to exploit high-efficient optical pumping of defect spins in wide-bandgap materials, such as silicon carbide (SiC) \cite{Kraus:2013di} and diamond \cite{Jin:2015eo}. Indeed, room-temperature stimulated emission due to population inversion of the silicon vacancy ($\mathrm{V_{Si}}$) spin states in SiC has been demonstrated \cite{Kraus:2013di}. However, MW amplification with positive gain in SiC has not been realized yet. Here, we demonstrate that the inversion factor---one of the crucial parameters required to achieve positive MW gain---can be about $I \approx 75$, corresponding to the population inversion  $| \rho_{0} | = 3 \%$ or the effective negative temperature of $4 \, \mathrm{K}$. 
This value is achieved in an X-band MW cavity  at room temperature under optical pumping with saturation power $P_0 = 0.35 \, \mathrm{W}$. 

Furthermore, because the saturation pump power scales with the spin-lattice relaxation rate $P_0 \propto 1 / T_1$ \cite{Riedel:2012jq}, we examine $T_1$ and find that it is in the sub-ms range at room temperature \cite{Simin:2017iw} but can be extremely long $T_1 > 60 \, \mathrm{s}$ at cryogenic temperature. Under these conditions, a very high population inversion above $| \rho_{0} |> 30 \%$ 
is achieved already for a very low pump power $P_0 < 0.01 \, \mathrm{W}$. 

\section{Pump efficiency at room temperature}

\begin{figure}[t]
\includegraphics[width=.47\textwidth]{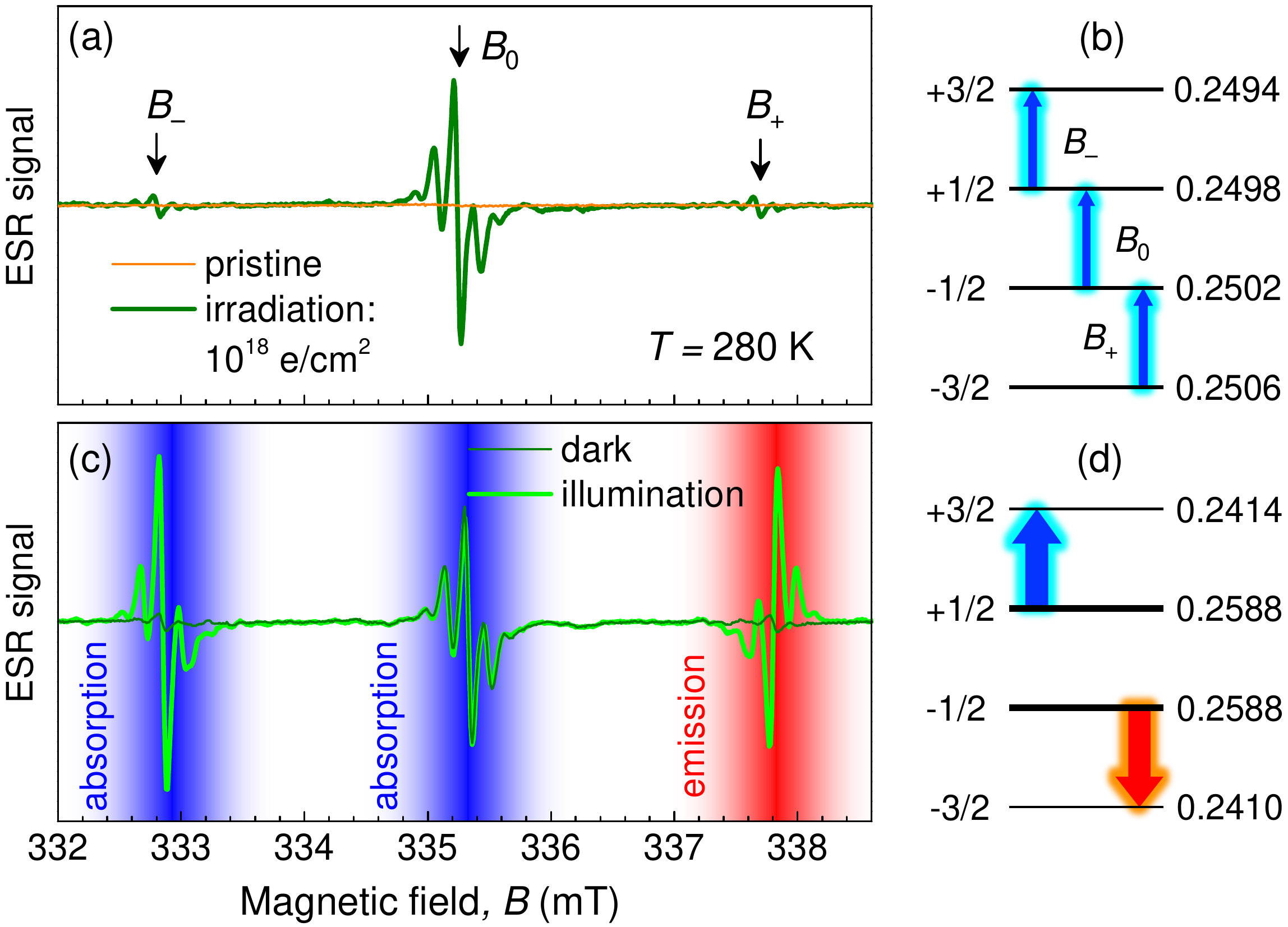}
\caption{Room temperature optical pumping of the  $\mathrm{V_{Si}}$ spins. (a) ESR spectra obtained in the dark from a pristine sample and after electron irradiation with a fluence of $1 \times 10 ^{18} \, \mathrm{cm}^{-2}$. (b) MW absorption between the $\mathrm{V_{Si}^-}$ spin sublevels in the dark due to the Boltzmann distribution. (c) ESR spectra obtained in the dark  and under illumination from an electron-irradiated sample. (d) MW absorption and stimulated emission between the $\mathrm{V_{Si}^-}$ spin sublevels under strong optical spin pumping ($P \gg P_0$) of the $m_s = \pm 1/2$ states. } \label{fig1}
\end{figure}

In all experiments reported in this Letter, we use a high-purity semi-insulating (HPSI) 4H-SiC wafer purchased from Norstel. A piece of this wafer is inserted into an X-band cavity with a resonance frequency $\nu = 9.4 \,  \mathrm{GHz}$. We perform electron spin resonance (ESR) experiments as described elsewhere \cite{Riedel:2012jq} and find that the pristine sample is ESR silent [thin line in Fig.~\ref{fig1}(a)], indicating a very low density of paramagnetic centers. To create spin-active defects, the wafer is irradiated with 2-MeV electrons to a fluence of $1 \times 10^{18} \, \mathrm{e / cm^{2}}$. The irradiation is performed at room temperature without consecutive annealing. Upon irradiation, three resonances appear in the ESR spectrum shown in Fig.~\ref{fig1}(a) (thick line). The origin of the central resonance at $B_0 = 335.3 \, \mathrm{mT}$ is discussed below. The magnetic field positions of the outer resonances follow $B_{\pm} = B_0 \pm D (3 \cos^2 \theta -1) / g_e \mu_B$ \cite{Pake:1962vm,Orlinski:2003dw,Kraus:2013vf}, were $g_e = 2.0$ is the electron g-factor, $\mu_B$ is the Bohr magneton and $\theta = 0^{\circ}$ is the angle between the applied magnetic field direction and the $c$-axis of 4H-SiC. From $B_+ - B_-$ we determine the zero-field splitting $2D = 70 \, \mathrm{MHz}$, which points at the negatively-charged $\mathrm{V_{Si}^-}$  defects in 4H-SiC \cite{Sorman:2000ij,Kraus:2013vf}. The $\mathrm{V_{Si}^-}$  concentration  $\mathcal{N} = 7 \times 10^{15} \, \mathrm{cm^{-3}}$ is found from the comparison of the photoluminescene (PL) intensity with a reference sample \cite{Fuchs:2015ii}. 

The  $\mathrm{V_{Si}^-}$ defect has a spin $S = 3/2$ ground state, which is split in four sublevels in an external magnetic field \cite{Mizuochi:2002kl,Kraus:2013di,Simin:2016cp}.  In thermodynamic equilibrium, the population distribution is given by the Boltzmann statistics  $\exp (- h \nu / k_B T)$, as shown in Fig.~\ref{fig1}(b), and the population difference between each two nearest sublevels is $\rho_B = 0.04 \%$ at room temperature (Appendix~\ref{ESR_calib}). Because the MW absorption at $B_+$ ($B_-$) is proportional to the population difference $\rho_{-3/2} - \rho_{-1/2}$ ($\rho_{+1/2} - \rho_{+3/2}$), the ESR signal can be calibrated (Appendix~\ref{ESR_calib}). 

\begin{figure}[t]
\includegraphics[width=.47\textwidth]{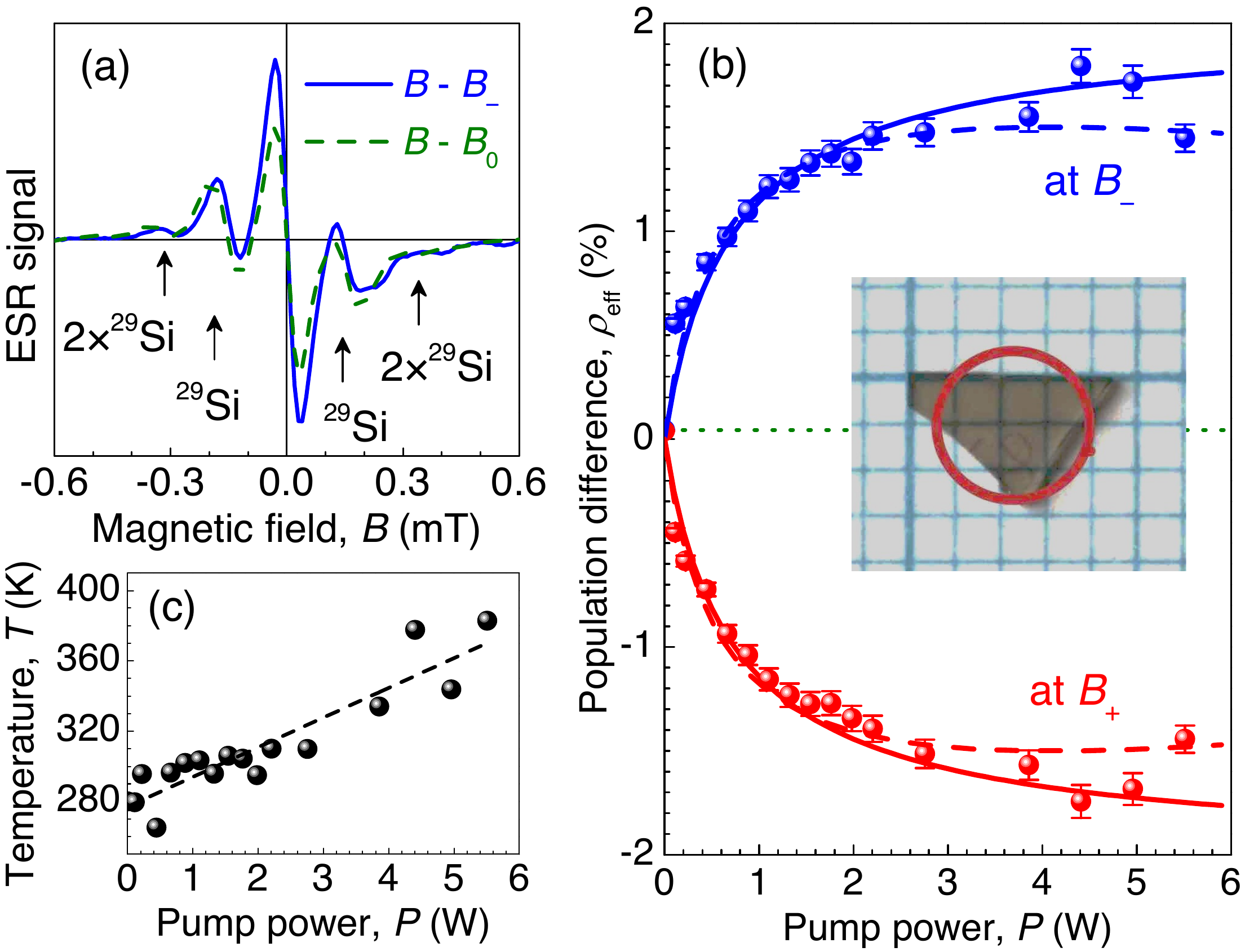}
\caption{Pump efficiency of the $\mathrm{V_{Si}^-}$ spins at room temperature. (a) Comparison of the central ESR peak at $B - B_0$ (dashed line) and the low-field peak at $B - B_-$ (solid line). The fine structure is due to the hyperfine interaction with one and two $^{29}$Si nuclear spins. (b) Optically induced population difference $\rho_{\mathrm{eff}}^{\pm}$ obtained from the ESR amplitude at $B_{\pm}$ as a function of pump power $P$. The pump wavelength is $\lambda = 805 \, \mathrm{nm}$. Negative signal corresponds to population inversion. The thin dotted line indicates difference $\rho_B = 0.04 \%$ owing to the the Boltzmann statistics at room temperature. The solid (dashed) lines are fits to Eq.~(\ref{Pol_Abs}) ignoring (assuming) sample heating. Inset:  A photo of the irradiated 4H-SiC crystal. The circle represents the laser spot (95\% of total intensity).  (c) Temperature of the 4H-SiC crystal as a function of $P$, obtained from the amplitude of the central ESR resonance. The dashed line is a linear fit. } \label{fig2}
\end{figure}

Figure~\ref{fig2}(a)  shows that the central (at $B_0$) and outer (at $B_{\pm}$) ESR resonances reveal the same fine structure due to the hyperfine interaction with one or two $^{29}$Si out of 12 silicon atoms in the next-nearest-neighbour (NNN) shell \cite{Vainer:1981vj,vonBardeleben:2000jg,Mizuochi:2003di}.  The central resonance is hence related to the silicon vacancy and it is suggested \cite{Mizuochi:2002kl,Janzen:2009ij,Soltamov:2015ez} to be negatively charged as well but with $D = 0$ due to higher $T_d$ symmetry compared to the $C_{3v}$ symmetry of $\mathrm{V_{Si}^-}$ \cite{Comment}. 

Under optical excitation with a wavelength $\lambda = 785 \, \mathrm{nm}$, the amplitude of the outer $\mathrm{V_{Si}^-}$ resonances drastically increases, while that of the central resonance remains the same as in the dark  [Fig.~\ref{fig1}(c)]. The most pronounced effect is observed at $B_+$ with the signal phase inversion, indicating that stimulated MW emission instead of MW absorption occurs \cite{Kraus:2013di}. Together with  $D > 0$ \cite{Orlinski:2003dw}, one concludes that the $m_s = \pm 1/2$ states are optically pumped and corresponding population probability can be as schematically shown in Fig.~\ref{fig1}(d). This property has been proposed for the realization of a room-temperature maser \cite{Kraus:2013di}, however the pump efficiency has not been investigated yet. 

\begin{figure}[t]
\includegraphics[width=.47\textwidth]{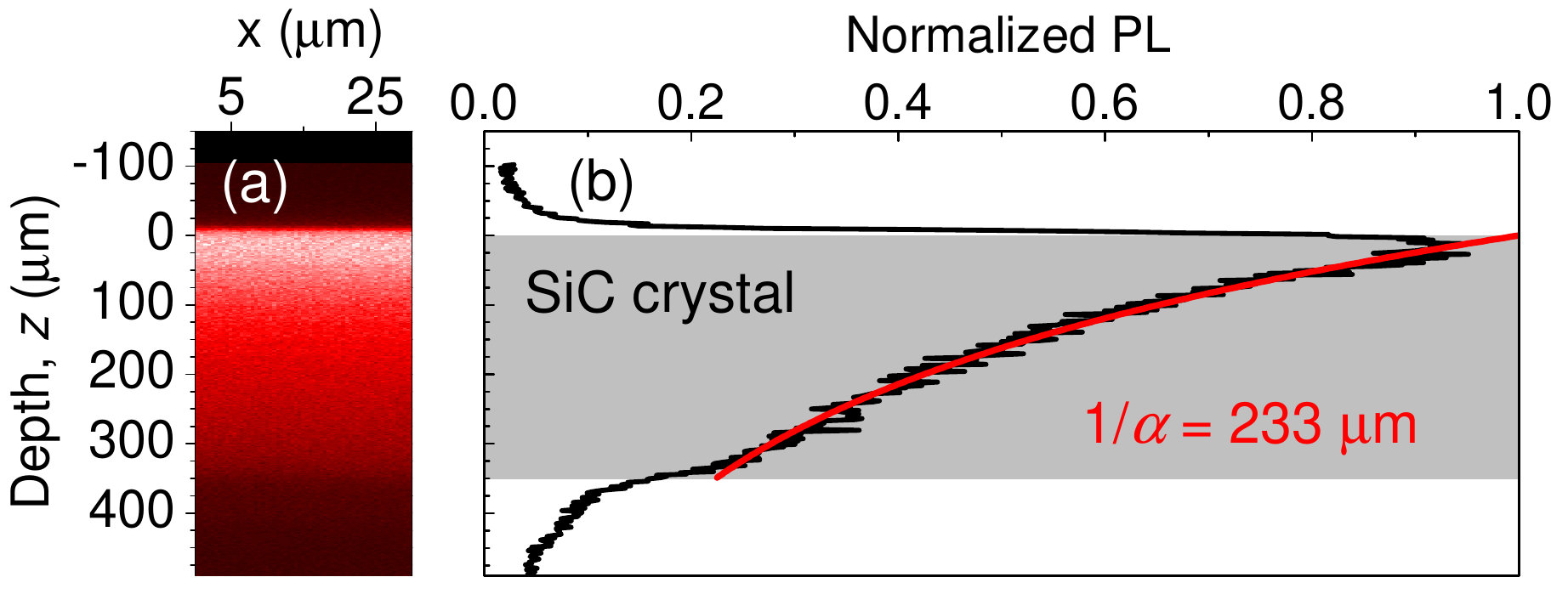}
\caption{Light absorption in SiC due the presence of $\mathrm{V_{Si}^-}$. (a) An $xz$-scan of the $\mathrm{V_{Si}}$ PL for the excitation wavelength of $805 \, \mathrm{nm}$. (b) Depth distribution of the PL intensity. The thick solid line is a fit to $\exp (- \alpha z)$. } \label{fig2_5}
\end{figure}

Using the thermal spin polarization of the central ESR line in Fig.~\ref{fig1}(a,b) as a reference (Appendix~\ref{ESR_calib}), we plot the effective spin polarization $\rho_{\mathrm{eff}}^{\pm}$, obtained at $B_{\pm}$ respectively, as a function of optical pump power $P$ in Fig.~\ref{fig2}(b). Here, $\rho_{\mathrm{eff}}^{+} < 0$ means population inversion. The polarization tends to saturate for  $P > 3 \, \mathrm{W}$. Note, that we have rotated the sample such that $\theta = 90^{\circ}$  (the external magnetic field is perpendicular to the $c$-axis) for better illumination conditions and we now use a high-power laser operating at a wavelength of $805 \, \mathrm{nm}$. Because the sample is opaque [inset of Fig.~\ref{fig2}(b)], the light penetration depth into the sample should be taken into account  to calculate the maximum achievable population difference  $\rho_0$. To do this, we perform a confocal $xz$-scan of the $\mathrm{V_{Si}^-}$ PL \cite{Kraus:2017cka} as presented in Fig.~\ref{fig2_5}(a). Fitting of the PL profile of Fig.~\ref{fig2_5}(b) to $\exp (- \alpha z)$ yields $1/ \alpha = 233 \, \mathrm{\mu m}$. Correspondingly, the absorption coefficient at the pump wavelength can be estimated as $1/  \alpha_{\mathrm{L}} > 1/ \alpha$ (Appendix~\ref{Absorption_coeff}). 

We now turn to the power dependence of the effective population difference, averaged over the sample thickness $\mathcal{L} = 350 \, \mathrm{\mu m}$, as a function of pump power $P$. Taking into account the absorption, it is described  for $| \rho_{\mathrm{eff}}^{\pm} | \gg \rho_B$ by (Appendix~\ref{PumpEff_abs})
\begin{equation}
\rho_{\mathrm{eff}}^{\pm}  (P) =  \mp \rho_{0} \frac{\phi}{\alpha_\mathrm{L} \mathcal{L}}  \, \ln \left(   \frac{P_0  + P}{P_0 + P \exp (-\alpha_\mathrm{L} \mathcal{L)}}  \right)    \,.
\label{Pol_Abs}
\end{equation}
Here, $\alpha_\mathrm{L} \mathcal{L} = 1.5$ is estimated from the experimental data of Fig.~\ref{fig2_5}(b). The sample is roughly triangle-shaped and approximately $\phi = 80\%$ of its area ($8 \, \mathrm{mm^2}$) is illuminated [inset of Fig.~\ref{fig2}(b)]. Eq.~(\ref{Pol_Abs}) fits the power dependencies of Fig.~\ref{fig2}(b) well for $P < 3 \, \mathrm{W}$  (solid lines). The two fitting parameters, the saturation population difference $\rho_0$ and characteristic power $P_0$, are summarized in table~\ref{AllParam} (see also Appendix~\ref{PumpEff_abs}). 

\begin{table}[t] 
\caption{Parameters $\rho_0$ and $P_0$ obtained from the fit of the experimental data to Eq.~(\ref{Pol_Abs}).  The laser spot area (95\% of the total intensity) is $9 \, \mathrm{mm^2}$. }
\begin{center}
\begin{tabular}{|c|c|c|}
Pump conditions & $\, \, \, \rho_0 \, \, \,$ & $P_0$  \\ \hline 
ignoring heating, Fig.~\ref{fig2}(b) & 2.5\% & $350 \, \mathrm{mW}$    \\ 
assuming heating, Fig.~\ref{fig2}(b) & 3\% & $P_0 \propto 1/T_1$    \\ 
low temperature, Fig.~\ref{fig3}(c) &  $>30\%$    &  $0.4 \, \mathrm{mW}$    \\ 
\end{tabular}
\end{center}
\label{AllParam}
\end{table}

The decrease of $| \rho_{\mathrm{eff}}^{\pm} |$ for $P > 3 \, \mathrm{W}$ is ascribed to sample heating. To measure the actual temperature, we use the central ESR resonance at $B_0$, whose amplitude should follow the Boltzmann statistics in the absence of optical pumping (Appendix~\ref{ESR_calib}).  The measured temperature as a function of pump power is presented in Fig.~\ref{fig2}(c) (symbols) and a linear fit to 
\begin{equation}
T (P) = T_0 + \xi P 
\label{Temp_P}
\end{equation}
gives $T_0 = 280 \, \mathrm{K}$ and $\xi = 17 \, \mathrm{K W^{-1}}$ (dashed line). The characteristic pump power is proportional to the spin-lattice relaxation rate $P_0 \propto 1 / T_1$ \cite{Riedel:2012jq} (see also Appendix~\ref{PumpEff_abs}), which in turn increases with temperature as $1/T_1 = A_5 T^5$ with $A_5 = 10^{-9} \, \mathrm{K^{-5} s^{-1}}$  \cite{Simin:2017iw}. Combining all together, we obtain $P_0$ with increasing pump power and then recalculate $\rho_{\mathrm{eff}}^{\pm}$ using Eq.~(\ref{Pol_Abs}) with parameters from table~\ref{AllParam}. The result of this calculation, shown by the dashed lines in Fig.~\ref{fig2}(b), describes well the non-monotonic behavior of the photo-induced population difference. Remarkably, the obtained $\rho_0 = 3\%$ corresponds to the Boltzmann distribution at a negative temperature of $4 \, \mathrm{K}$, and the corresponding inversion ratio $I = \rho_0 / \rho_{B}$ can be very high approaching $I = 75$.

\section{Spin-lattice relaxation time at cryogenic temperature}

\begin{figure}[t]
\includegraphics[width=.47\textwidth]{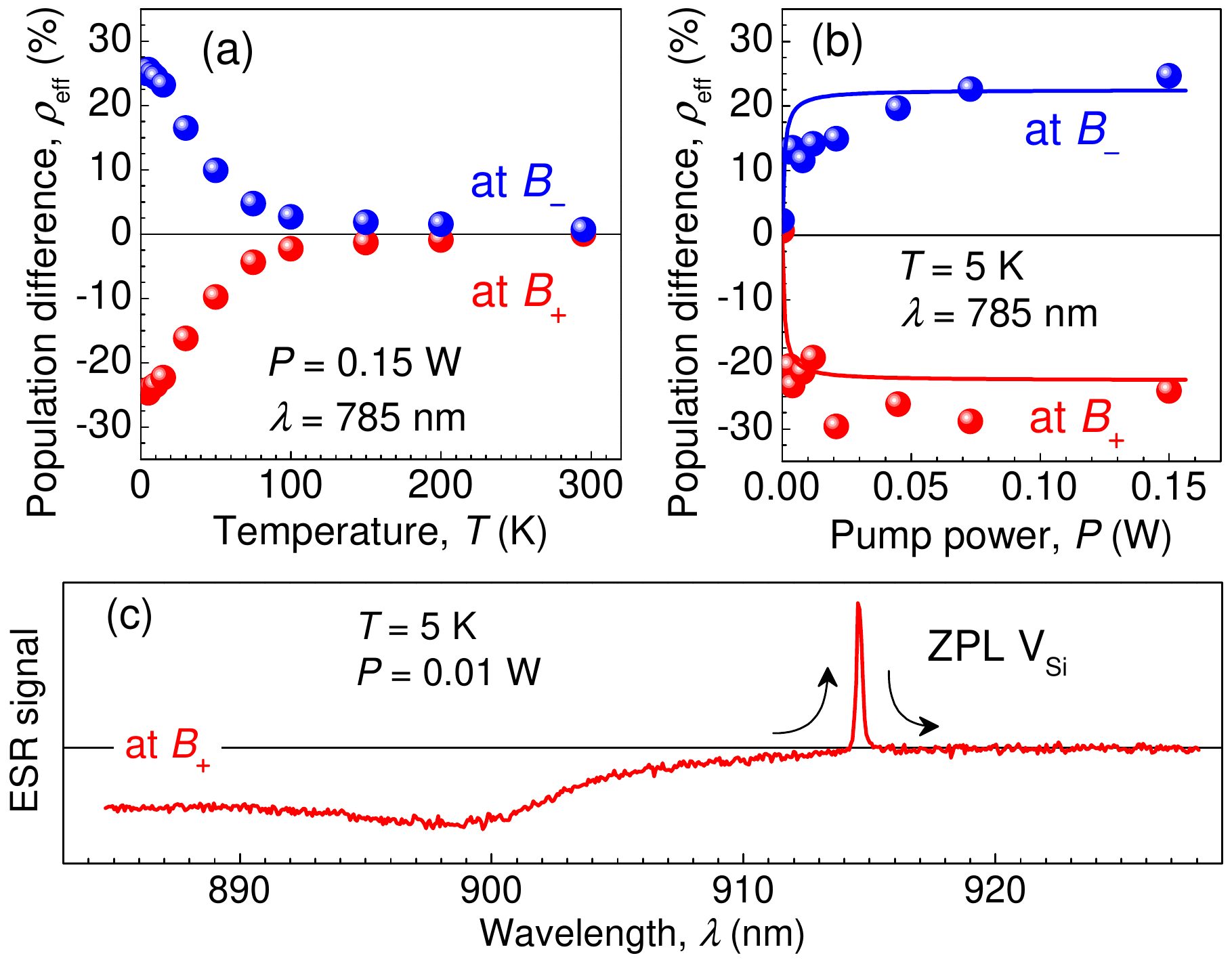}
\caption{Pump efficiency of the $\mathrm{V_{Si}^-}$ spins at cryogenic temperature $T = 5 \, \mathrm{K}$. (a) Optically induced population difference $\rho_{\mathrm{eff}}^{\pm}$ obtained from the ESR amplitude at $B_{\pm}$ as a function of temperature. The pump wavelength is $\lambda = 785 \, \mathrm{nm}$. (b) Optically induced population difference as a function of pump power. The solid lines are fits to Eq.~(\ref{Pol_Abs}). (c) 
ESR signal at $B_+$ as a function of the pump wavelength. The peak at $\lambda_{\mathrm{ZPL}} = 914.5 \, \mathrm{nm}$ corresponds to the $\mathrm{V_{Si}^-}$ ZPL.  The pump power is $P = 0.01 \, \mathrm{W}$. The arrows indicate the scan direction, from shorter to longer wavelengths. } \label{fig3}
\end{figure}

As it follows from the previous section, reduction of the spin-lattice relaxation rate results in a reduction of the characteristic pump power. Figure~\ref{fig3}(a) presents the effective population difference as a function of temperature at a fixed pump power  $P = 150 \, \mathrm{mW}$.  There is a drastic increase of $| \rho_{\mathrm{eff}}^{\pm} | $ for $T < 100 \, \mathrm{K}$, which mirrors the temperature dependence of $T_1$ \cite{Simin:2017iw}. At $T = 5 \, \mathrm{K}$, a larger population difference $| \rho_{\mathrm{eff}}^{\pm} |  > 25 \%$ is achieved already at a very low pump power $P < 1 \, \mathrm{mW}$ [Fig.~\ref{fig3}(b) and table~\ref{AllParam}]. The real polarization is even higher because in case of long $T_1$, even a very weak MW field reduces $| \rho_{\mathrm{eff}}^{\pm} |$ effectively through absorption and stimulated emission processes.  

To probe spin-lattice dynamics at low temperature, we first measure the ESR signal at $B_+$ as a function of pump wavelength $\lambda$, shown in Fig.~\ref{fig3}(c). The optimum wavelength to achieve the population inversion is $\lambda = 900 \, \mathrm{nm}$ and optical pumping becomes inefficient for $\lambda > 910 \, \mathrm{nm}$. In addition to the main trend, there is a sharp peak at $\lambda_{\mathrm{ZPL}} = 914.5 \, \mathrm{nm}$, corresponding to the resonant excitation into the $\mathrm{V_{Si}^-}$ zero-phonon line (or in alternative notations the V2 ZPL) \cite{Sorman:2000ij,Riedel:2012jq}. Interestingly, under resonance excitation, MW absorption rather than stimulated emission occurs, indicating that the pump cycle might be different, i.e., pumping the $m_S = \pm 3/2$ rather than the $m_S = \pm 1/2$ states. At $B_-$ the behavior is exactly opposite. Detailed discussion of this effect is beyond the scope of this work.  

\begin{figure}[t]
\includegraphics[width=.45\textwidth]{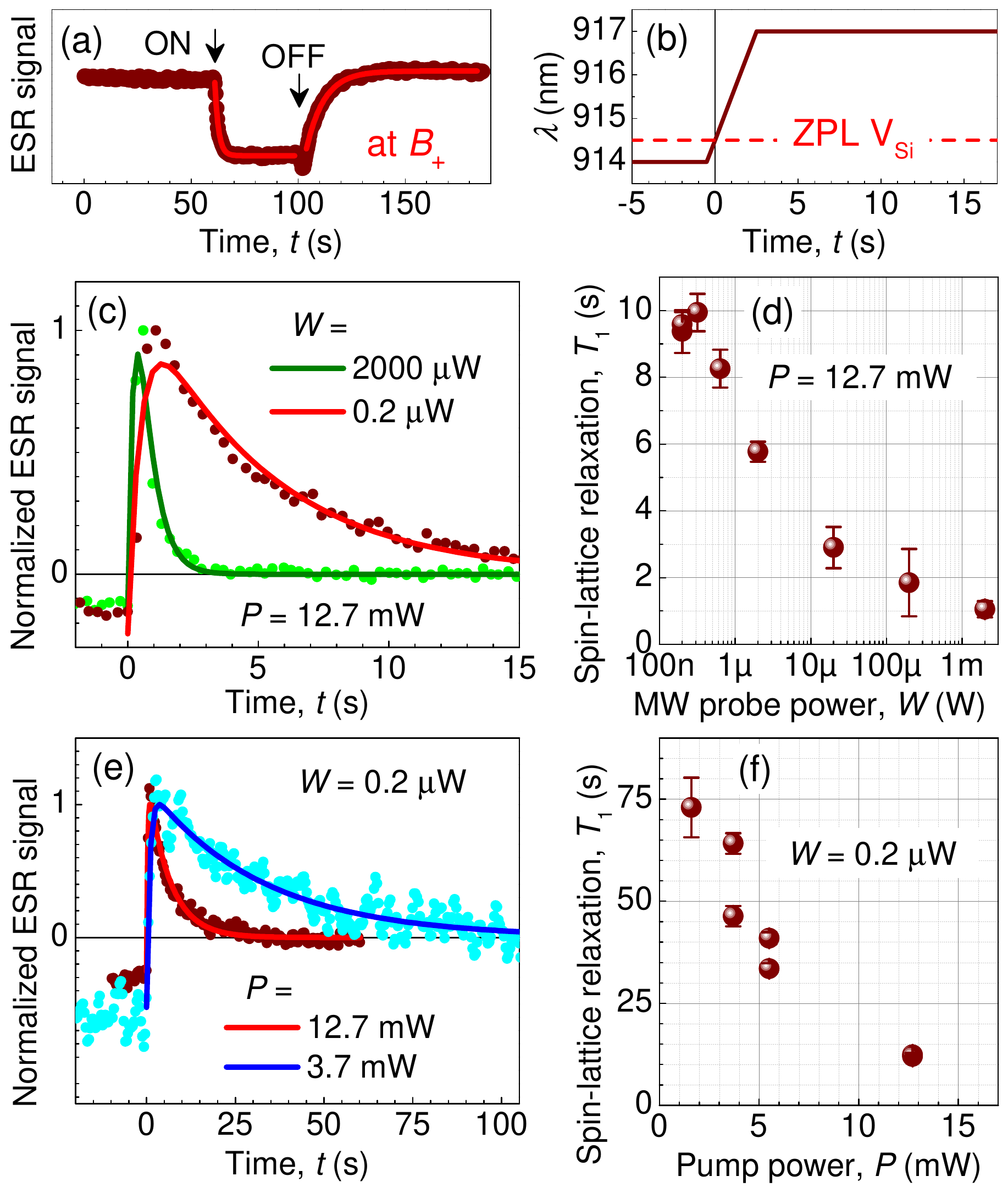}
\caption{Spin-lattice relaxation time at $T = 3.6 \, \mathrm{K}$. (a) Transient ESR signal at $B_-$ on the laser ($\lambda = 900 \, \mathrm{nm}$) on-off cycling. The solid lines are mono-exponential fits.  (b) $T_1$-measurement protocol. The laser wavelength is swept through the $\mathrm{V_{Si}^-}$ ZPL (from $914$ to $917 \, \mathrm{nm}$) with simultaneous monitoring of the transient ESR signal. (c) ESR signal decay for different MW probe power $W$. The solid lines are two-exponential fits to Eq.~(\ref{Two_Exp}). (d) The spin-lattice relaxation time $T_1$ as a function of $W$. (e) ESR signal decay for different optical pump power $P$. The solid lines are two-exponential fits to Eq.~(\ref{Two_Exp}). (f) The spin-lattice relaxation time $T_1$ as a function of $P$.} \label{fig4}
\end{figure}

Second, we use the simplest procedure, namely we switch the non-resonant excitation on and off, while monitoring the ESR signal as a function of time [Fig.~\ref{fig4}(a)]. The transient ESR signal is governed by two processes: (i) spin-lattice relaxation and (ii) restoring/changing of the sample temperature and cavity characteristics under illumination. To keep the cavity and sample heating under the same condition, i.e. to exclude contribution (ii), we apply another protocol, shown in Fig.~\ref{fig4}(b). Here, the laser is always switched on, but its wavelength is scanned through the ZPL, i.e. the spin pumping takes place only when the wavelength exactly matches $\lambda_{\mathrm{ZPL}} = 914.5 \, \mathrm{nm}$ for a few ms during the scan.  

The corresponding transient ESR signal $S_{\mathrm{ESR}}$ is presented in Fig.~\ref{fig4}(c). To compare different MW powers $W$,  the signal is normalized and the time scale is shifted such that at $t = 0$ the pump wavelength is equal to the ZPL.  The experimental data for $t  > 0 $ can then be well fitted to 
\begin{equation}
S_{\mathrm{ESR}} = -A_P  \exp (-t/\tau_P)  + A_1 \exp (- 2t / T_1) \,.
\label{Two_Exp}
\end{equation}
Here, the rise time $\tau_P$ reflects the pump rate and instrument response function, while the decay time corresponds to $T_1$ \cite{Widmann:2014ve}. Remarkably, $A_1 - A_P < 0$ indicates that at $\lambda < \lambda_{\mathrm{ZPL}}$ ($t < 0$) the optical pump is weak but not negligable, in accord to Fig.~\ref{fig3}(c). Figure~\ref{fig4}(d) summarizes $T_1$ when the MW probe power varies over 4 orders of magnitude and for $W = 200 \, \mathrm{nW}$ we obtain $T_1 = 9.4 \pm 0.6 \, \mathrm{s}$. It agrees with the previously reported $T_1$, obtained  from another pulsed measurement protocol \cite{Simin:2017iw}.  

To minimize any possible effect of illumination and to further increase $T_1$, we repeat measurements for the lowest MW probe power of $200 \, \mathrm{nW}$ while reducing the optical pump power [Fig.~\ref{fig4}(e)].  Again from fits to Eq.~(\ref{Two_Exp}), we find $T_1$. As expected, it increases with decreasing $P$ [Fig.~\ref{fig4}(f)] and for $P = 1.6 \, \mathrm{mW}$ we achieve an exceptionally long spin-lattice relaxation time $T_1 = 73 \pm 7 \, \mathrm{s}$. This value is comparable to that reported for the NV defects in diamond in the mK temperature range \cite{Amsuss:2011ci,Ranjan:2013cr}.

\section{Discussion}

\subsection{Room-temperature continuous-wave maser}

Based on the above results, it is conceivable to implement a maser amplifier with the $\mathrm{V_{Si}^-}$ centers. We present two different approaches. First off, we make some considerations with respect to the magnetic quality factor  $Q_m$. For best maser performance, one needs the lowest possible $Q_m$ or the highest reciprocal Q-factor $1/Q_m$, which is given by \cite{siegman1964microwave} 
\begin{equation}
\frac{1}{Q_m} = 2 K \frac{h \nu}{k_B T} \frac{\mathcal{N}}{n} \frac{I \beta^2 \eta}{\Delta \nu}  = -  2 K  \,  \frac{\beta^2 \eta}{\Delta \nu} \mathcal{N} \rho_{\mathrm{eff}}\,.
\label{Qm}
\end{equation}
Here, $K = g_e^2 \mu_B^2 \mu_0 / h = 0.65 \times 10^{-18} \, \mathrm{m^3 Hz}$ (or $\mathrm{cm^3 MHz}$) and $n = 4$ is the number of spin sublevels.  For the $(-1/2 \rightarrow -3/2)$ transition (at $B_+$) the matrix element is $\beta^2 = 3/4$ \cite{siegman1964microwave} . The corresponding resonance linewidth $\Delta \nu = 1.4 \, \mathrm{MHz}$ is obtained from Fig.~\ref{fig1}(c). With these experimental parameters and for $\rho_{\mathrm{eff}}^+ = -1.5\%$, we have $Q_m = 1.4 \times 10^4 / \eta$. The masing effect occurs when the Q-factor of the maser cavity $Q_0 > Q_m$. Thus, it is necessary to use high-Q cavities and simultaneously achieve a high filling factor  $\eta \rightarrow 1$. 

\begin{figure}[t]
\includegraphics[width=.48\textwidth]{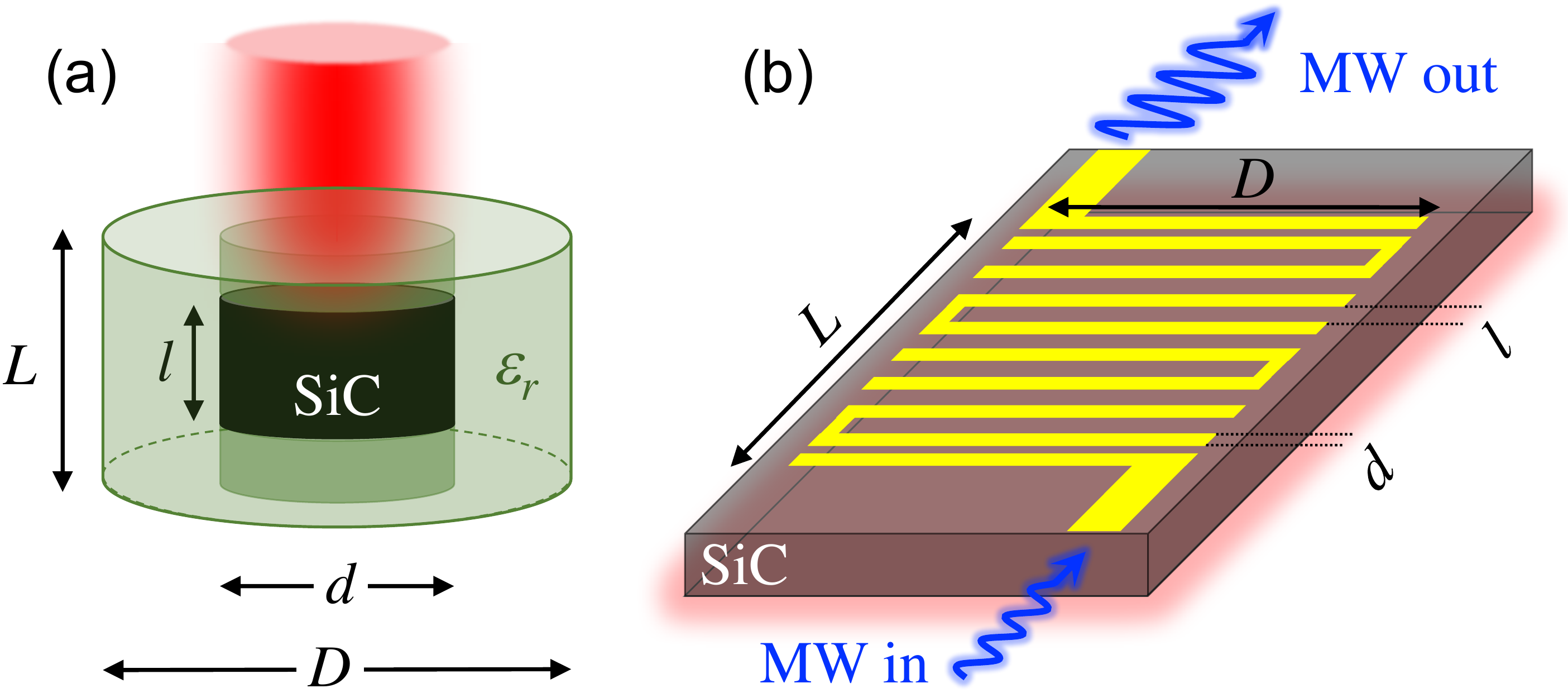}
\caption{SiC maser amplifiers. (a) A scheme of a dielectric cylindrical cavity, the inner part is filled with SiC. The cavity is placed in a metal can (not shown), and the optical pump is performed through the top window. (b) An optically pumped  
hairpin bandpass filter deposited on a SiC substrate of $350 \, \mathrm{\mu m}$ thickness. The MW signal is amplified due to stimulated emission in SiC. } \label{figS}
\end{figure}

For this purpose, we now consider a cylindrical cavity consisting of a $\epsilon_r = 10$ puck with diameter $D = 8 \, \mathrm{mm}$ and height $L = 4  \, \mathrm{mm}$ [Fig.~\ref{figS}(a)]. The $\mathrm{TE}_{01 \delta}$ mode resonance has a frequency of $12 \, \mathrm{GHz}$ \cite{Breeze:2016da}. According to the simulation \cite{Breeze:2016da}, more than $\eta > 80 \%$ of the magnetic energy is concentrated in the middle of the puck with $d \approx D/2$ and $l \approx L/2$ [Fig.~\ref{figS}(a)]. It is natural to use sapphire with low dielectric losses for the puck and fill the inner part with SiC, both materials having dielectric constants $\epsilon_r $ close to 10. With the help of  Eq.~(\ref{Pol_Abs}), the required pump power for such a maser can be estimated as $P_m \approx 2 \pi (d/2)^2 \tilde{P_0} e^{\alpha_{\mathrm{L}} l}$. Here, $\tilde{P}_0  = 3.9 \, \mathrm{W / cm^{2}}$ is the characteristic pump power per area (table~\ref{AllParam})  and we obtain $P_m = 5.3 \, \mathrm{kW}$. Due to unavoidable heating, such a high pumping regime can likely be realized in pulsed mode only. 

To  decrease the optical pump power one can reduce the absorption coefficient $\alpha_{\mathrm{L}}$. We use the Beer-Lambert law $\alpha_{\mathrm{L}} = \sigma \mathcal{N}$, where $\sigma = 6.1 \times 10^{-15} \, \mathrm{cm^{2}}$ (according to Fig.~\ref{fig2_5}) is the absorption cross-section  averaged over all types of irradiation defects. For a defect density lowered to $\mathcal{N} = 2 \times 10^{15} \, \mathrm{cm^{-3}}$, we estimate the absorption coefficient $1/ \alpha_{\mathrm{L}} \gtrsim800 \, \mathrm{\mu m}$ and the required pump power to achieve $\rho_{\mathrm{eff}}^+ = -1.5\%$ is $P_m = 11 \, \mathrm{W}$ or $\tilde{P}_m= 21 \, \mathrm{W / cm^{2}}$. Using experimental data of Fig.~\ref{fig2}(b), we estimate that the effect of heating (i.e., deviation of the dashed and solid lines) becomes significant for $\tilde{P}> 30 \, \mathrm{W / cm^{2}}$. This means that cw operation mode of the maser under the aforementioned conditions is feasible.  Remarkably, we then obtain $Q_m = 6 \times 10^4$ from Eq.~(\ref{Qm}) and $Q_0 = 2 \times 10^5  > Q_m$ has been reported for a sapphire cylindrical cavity at room temperature \cite{Oxborrow:2012dd}. 

Further improvement (i.e.,  pump power reduction, decreasing of $Q_m$, smaller cavity volume) can be achieved using dielectric cavities with high $\epsilon_r$ \cite{Breeze:2015ea}. In this case, dielectric losses can be significant and exact modelling is required, which is beyond the scope of this work. Alternatively, the cavity dimensions $L$ and $D$ are smaller for the K-band (18 to 27~GHz) and Ka-band (27 to 40~GHz) compared to the X-band (8 to 12~GHz), which we considered so far. This means that the pre-factor $e^{\alpha_{\mathrm{L}} l}$ is the same as in the above consideration for higher $\alpha_{\mathrm{L}}$ and hence higher $\mathcal{N}$, resulting in lower $Q_m$ and better maser performance at higher frequencies. 


\subsection{Cryogenic maser amplifier}

At cryogenic temperature, magnetic quality factor $Q_m$ can be significantly lower because of higher pump efficiency and higher photo-induced population difference (table~\ref{AllParam} ). For the cylindrical cavity of Fig.~\ref{figS}(a) with $\mathcal{N} = 1 \times 10^{17} \, \mathrm{cm^{-3}}$ under illumination with $P = 600 \, \mathrm{mW}$, one obtains $Q_m = 70$ using Eq.~(\ref{Qm}). Thus, cw masing should be achievable even under sub-Watt pumping using commercial cavities with $Q_0 \sim 10^3 - 10^4$.   

Alternatively, we now simulate---with the help of \textit{Sonnet Software}---a traveling-wave maser amplifier based on a superconducting hairpin bandpass filter deposited on a 350-$\mathrm{\mu m}$-thick SiC substrate [Fig.~\ref{figS}(b)]. The amplitude gain coefficient is $\alpha_m = 2 \pi \nu /Q_m v_g$, where $v_g$ is the group velocity \cite{siegman1964microwave}. The maser power gain in decibels can then be expressed as 
\begin{equation}
G_{\mathrm{dB}} \approx 27 \frac{\mathcal{S} \Lambda}{Q_m}   \,.
\label{Gain}
\end{equation}
Here, $\Lambda = L / \lambda$ is the length of the circuit ($L \sim 9 \, \mathrm{mm}$) relative to the MW wavelength in vacuum ($\lambda$). For the resonance frequency $\nu = 10.4 \, \mathrm{GHz}$ (defined through the length $D \sim 5 \, \mathrm{mm}$ of the hairpin) one has $\lambda = 29 \, \mathrm{mm}$ and, correspondingly, $\Lambda = 0.3$. The slowing factor $\mathcal{S} = c / v_g$ depends on the geometry of the structure.  For the parameters $l \sim 0.8 \, \mathrm{mm}$ and $d \sim 0.3 \, \mathrm{mm}$ given in Fig.~\ref{figS}(b), the simulation at $\nu = 10.4 \, \mathrm{GHz}$ gives $\mathcal{S} = 80$.  We assume $Q_m = 70$ as in the previous example and obtain $G_{\mathrm{dB}} = 9 \, \mathrm{dB}$ for pump power $P = 2.3 \, \mathrm{W}$. For applications as low-noise MW amplifiers, it is necessary to increase the gain, and $G_{\mathrm{dB}} \approx 30 \, \mathrm{dB}$ can be achieved by enhancement of the circuit length to $\Lambda = 1$ and by simultaneous enhancement of the pump power to $P = 7.7 \, \mathrm{W}$. Furthermore, lowering of $P$ is possible by (i) reducing area of the circuit using coplanar waveguides \cite{Schuster:2010bu, Kubo:2010iq} and (ii) creating $\mathrm{V_{Si}^-}$ defects only below the feedline where the MW field is maximal (spatial control over the $\mathrm{V_{Si}^-}$ creation can be realized with various methods \cite{Falk:2013jq, Kraus:2017cka, Wang:2016vn}). 

\section{Summary}

We measured the pump efficiency of the $\mathrm{V_{Si}^-}$ defects in 4H-SiC under optical illumination. The population difference reaches $3\%$ at room temperature, which corresponds to the population inversion with the effective negative temperature of $4 \, \mathrm{K}$. Based on the experimentally obtained parameters, a design of a room temperature maser operating in continuous-wave mode and its pump conditions are discussed. The pump efficiency increases by several orders of magnitude by lowering the bath temperature and the population inversion can be above $30\%$.  This is a consequence of an exceptionally long spin-lattice relaxation time exceeding one minute. Besides cryogenic receivers \cite{Narkowicz:2013fb} and ultralow-noise quantum microwave amplifiers, a coherent coupling of optically pumped $\mathrm{V_{Si}}$ spins to planar (high-$T_c$) superconducting cavities can be realized \cite{Schuster:2010bu,Kubo:2010iq,Wu:2010fm,Ghirri:2016hw}. A single-spin maser, where many indistinguishable MW photons are generated by one and the same $\mathrm{V_{Si}^-}$ center in a superconducting cavity \cite{Astafiev:2007gd}, and a strong coupling of a single spin to a MW photon \cite{Mi:2017ex} are intriguing perspectives. Our findings hence suggest that SiC with optically active spin defects is a promising platform for microwave photonics and quantum electronics. 

\begin{acknowledgments}
We gratefully thank J.~Fichtner, D.~Poprygin, C.~Kasper for test measurements and M.~Auth for technical help. This work has been supported by the German Research Foundation (DFG) under grants DY 18/13 and AS 310/5 as well as by KAKENHI(A) 17H01056. 
\end{acknowledgments}


\appendix

\section{Calibration of the ESR signal} \label{ESR_calib}

Without saturation, the MW absorption and therefore the ESR peak-to-peak (P2P) signal at $B_{\pm}$ is proportional to the population difference, namely $\Delta \rho ^{+} = \rho_{-3/2} - \rho_{-1/2}$ and $\Delta \rho ^{-} = \rho_{+1/2} - \rho_{+3/2}$, as   
\begin{equation}
S_{\mathrm{ESR}}^{\pm} = \gamma  \, \Delta \rho^{\pm}  \, \mathcal{N} \,. 
\label{P2P}
\end{equation}
Here, $ \gamma$ is the proportionality constant, which depends on the experimental conditions. The occupation probabilities of the spin states in the dark are given by the Boltzmann statistics 
\begin{align}
\rho_{-1/2} &= \rho_{-3/2} \, \exp (- h \nu / k_B T) \\
\rho_{+1/2} &= \rho_{-1/2} \, \exp (- h \nu / k_B T) \\
\rho_{+3/2} &= \rho_{+1/2} \, \exp (- h \nu / k_B T) \,.
\label{Boltz}
\end{align}
We neglect the influence of the zero field splitting as it is much smaller than the Zeeman splitting. With the MW frequency $\nu = 9.4 \, \mathrm{GHz}$ at a temperature $T_0 = 280K$ and using 
\begin{equation}
\rho_{-3/2} + \rho_{-1/2} +\rho_{+1/2} +\rho_{+3/2} = \sum \rho_{m_S} = 1 \,, 
\label{4Sum}
\end{equation}
we obtain $\rho_{m_S}$ as shown in Fig.~\ref{fig1}(b). We then compare these calculations with experimental data of Fig.~\ref{fig1}(a) and with help of  Eq.~(\ref{P2P}) obtain $\gamma \mathcal{N}$. To reconstruct the occupation numbers under illumination, the procedure is reversed. First, we use the calibrated ESR signal to obtain the population difference $\Delta \rho^{\pm} = S_{\mathrm{ESR}}^{\pm}  / \gamma \mathcal{N}$.  Then, with $\rho_{-1/2} = \rho_{+1/2}$ and $\sum \rho_{m_S} = 1$, we calculate all four $\rho_{m_S}$ as shown in Fig.~\ref{fig1}(d). 

The central ESR signal originates from another $V_{\mathrm{Si}}$ configuration, but it can still be approximated for $h \nu \gg k_B T$ to
\begin{equation}
S_{\mathrm{ESR}}^{0}  \propto  \rho_B \approx \frac{h \nu}{n k_B T} \,, 
\label{P2P_B0}
\end{equation}
with $n$ being the number of spin sublevels. Because $S_{\mathrm{ESR}}^{0}$ is not influenced by spin pumping, we can perform a back calculation on the sample temperature as a function of pump power [Fig.~\ref{fig2}(c)]
\begin{equation}
T(P) =   \frac{S_{\mathrm{ESR}}^{0} (0)}{S_{\mathrm{ESR}}^{0} (P)} T_0  \,. 
\label{P2P_Temp}
\end{equation}
Here, $S_{\mathrm{ESR}}^{0} (0)$ and $S_{\mathrm{ESR}}^{0} (P)$ denote the P2P signal at $B_0$ in the dark and under illumination power $P$, respectively. 

\section{Estimation of the absorption coefficient} \label{Absorption_coeff}

The  normalited PL intensity $I_{\mathrm{PL}}$  of Fig.~\ref{fig2_5}(b) follows 
\begin{equation}
I_{\mathrm{PL}} (z) =   \exp(- \alpha_{\mathrm{L}} z) \exp(- \alpha_{\mathrm{PL}} z) \,,  
\label{PL_profile}
\end{equation}
where $\alpha_{\mathrm{L}}$ and $\alpha_{\mathrm{PL}}$ are the absorption coefficients at the pump ($785 \, \mathrm{nm}$ or $805 \, \mathrm{nm}$) and PL (maximum at $910 \, \mathrm{nm}$) wavelengths, respectively. The fit to $\exp(- \alpha z)$ in Fig.~\ref{fig2_5}(b) yields, therefore, $\alpha = \alpha_{\mathrm{L}}  + \alpha_{\mathrm{PL}}$. Given that the absorption coefficient increases with decreasing wavelength  \cite{Wendler:2012co,Hain:2014tl}, i.e., $\alpha_{\mathrm{L}}  > \alpha_{\mathrm{PL}}$, the absorption at the pump wavelength is restricted to $\alpha / 2 < \alpha_{\mathrm{L}} <  \alpha$.  To estimate the pump power, we use the least favourable scenario $\alpha_{\mathrm{L}} \approx  \alpha$, indicating that the characteristic pump power $P_0$ can be even lower than estimated in the main text. 

\section{Effect of light absorption on pump efficiency} \label{PumpEff_abs}

We consider two spin sublevels $m_S = +1/2$ and $m_S = + 3/ 2$ with corresponding population probabilities $\rho_{+1/2}$ and $\rho_{+3/2}$ [Fig.~\ref{fig1}(d)].  The population difference between the lower and higher energy sublevels $\Delta \rho ^{-} = \rho_{+1/2} - \rho_{+3/2}$ is defined through two processes: (i) optical spin pumping $\kappa P (\rho_0 - \Delta \rho)$ and (ii) spin-lattice relaxation $(\rho_\mathrm{B} - \Delta \rho) / T_1$. Here, $\kappa$ is a coefficient of proportionality between the spin pumping rate and  the pump power $P$. In steady-state, these processes compensate each other yielding 
\begin{equation}
 \Delta \rho^{-} =  \frac{P}{P + P_0} \rho_0 +  \frac{P_0}{P + P_0} \rho_B \,,
\label{Drho_steady_minus}
\end{equation}
with $P_0 = 1 / \kappa T_1$ being a characteristic pump power. 

In case of another two spin sublevels $m_S = -1/2$ and $m_S = - 3/ 2$ with $\Delta \rho ^{+} = \rho_{-3/2} - \rho_{-1/2}$, a population inversion is optically generated $\Delta \rho ^{+} < 0$ [Fig.~\ref{fig1}(d)]. We assume the same maximum achievable population difference $\rho_0$ under  optical spin pumping and obtain 
\begin{equation}
 \Delta \rho^{+} =  - \frac{P}{P + P_0} \rho_0 +  \frac{P_0}{P + P_0} \rho_B \,.
\label{Drho_steady_plus}
\end{equation}
For $P \gg P_0 \rho_B / \rho_0$, the population difference can be approximated to 
\begin{equation}
 \Delta \rho^{\pm} =  \mp \frac{P}{P + P_0} \rho_0 \,.  
\label{Drho_steady_P}
\end{equation}
Due to the laser light absorption, the pump power decreases with depth as $P \exp (- \alpha z)$ and to obtain the effective population difference, it is necessary to integrate over the sample thickness 
\begin{equation}
\rho_{\mathrm{eff}}^{\pm}  (P) =  \mp \rho_0  \frac{1}{\mathcal{L}}  \int_{0}^{\mathcal{L}}  \frac{P \exp(- \alpha_{\mathrm{L}} z)}{P \exp(- \alpha_{\mathrm{L}} z) + P_0}  dz   \,.  
\label{Drho_Intagrate_L}
\end{equation}
This integral can be solved analytically. Considering that only a fraction of the sample is illuminated [inset of Fig.~\ref{fig2}(b)]---described by the parameter $\phi$---one obtains  Eq.~(\ref{Pol_Abs}).




%

\end{document}